# Change in the magnetic structure of (Bi,Sm)FeO$_3$ thin films at the morphotropic phase boundary probed by neutron diffraction

Shingo Maruyama,[1)] Varatharajan Anbusathaiah,[1)] Amy Fennell,[2)] Mechthild Enderle,[3)] Ichiro Takeuchi,[1)] and William D. Ratcliff[4,a)]

1*Department of Materials Science and Engineering, University of Maryland, College Park, Maryland 20742, USA*

2*Paul Scherrer Institut, 5232 Villigen PSI, Switzerland*

3*Institut Laue Langevin, BP 156, 38042 Grenoble, France*

4*NIST Center for Neutron Research, National Institute of Standards and Technology, Gaithersburg, Maryland 20899,*

USA

We report on the evolution of the magnetic structure of BiFeO$_3$ thin films grown on SrTiO$_3$ substrates as a function of Sm doping. We determined the magnetic structure using neutron diffraction. We found that as Sm increases, the magnetic structure evolves from a cycloid to a G-type antiferromagnet at the morphotropic phase boundary, where there is a large piezoelectric response due to an electric-field induced structural transition. The occurrence of the magnetic structural transition at the morphotropic phase boundary offers another route towards room temperature multiferroic devices.

As one of the only single-phase multiferroic materials with ferroelectric and magnetic transition temperatures well above room temperature, BiFeO$_3$ (BFO) has been extensively studied. It possesses a robust ferroelectric polarization which is closely tied to its rhombohedral structure, and microstructural properties of BFO thin films (such as the stress state, grain size and its orientation, etc.) can sensitively affect its local ferroelectric properties. Magnetoelectric coupling between the local ferroelectric polarization and magnetism inside BFO thin films can serve as the basis for heterostructured multiferroic devices,[1,2] but their antiferromagnetic properties are known to display complex variations depending delicately on the local microstructural properties. We have previously used neutron diffraction to probe the nature of antiferromagnetic domains in epitaxial BFO thin films.[3,4]

Chemical substitution in BFO has been explored in order to improve the ferroelectric, piezoelectric, and dielectric properties of the material.[5–8] It has been demonstrated that the substitution of rare earth elements into the A-site of BFO thin films results in a structural phase transition from a ferroelectric rhombohedral phase to a paraelectric orthorhombic phase.[9,10,11] In the case of Sm the transition occurs at ~14% doping[12] at which point films exhibit a Morphotropic Phase Boundary (MPB). Earlier studies showed that in the vicinity of the MPB, an electric field can be used to drive the transition from the paraelectric orthorhombic phase to the rhombohedral ferroelectric phase, resulting in a very large piezoelectric effect d$_{33}$

___________________________

a) Electronic mail: william.ratcliff@nist.gov

larger than 150 pm/V.[13] Since the ferroelectric and magnetic domains are coupled in this compound,[10,14–17] the change of the magnetic structure across the phase boundary can potentially serve as another avenue for a multiferroic device operation. In this study, we determined the magnetic structure of Sm-doped BFO films grown on a SrTiO$_3$ (STO) substrate using polarized neutron diffraction. We show that the magnetic structure is sensitive to Sm doping. For Sm concentration less than ~14 %, indications for a cycloid were found, while for Sm concentration above ~14 %, a G-type collinear magnetic order is observed.

BFO films grown on a (0 0 1) STO substrate show four equivalent crystallographic domains based on the four different quasi-cubic body diagonals, which complicates the determination of the magnetic structure. In order to simplify our investigation of the magnetic structure Sm-doped BFO films, we used (0 0 1) STO substrates with a 4° miscut along [1 1 0] direction on which we can obtain single ferroelectric domain BFO films.[18,19] The Sm-doped BFO films are grown epitaxially by Pulsed Laser Deposition (PLD) with alternating deposition of Bi-rich BFO and SmFeO$_3$ targets onto the ~50 nm SrRuO$_3$ (SRO) buffered miscut (0 0 1) STO substrates. The thickness of the Sm doped BFO films are ~840 nm for the 10.4% Sm-doped BFO film and ~1 μm for the 16.1 and 18.7% Sm-doped BFO films. The epitaxial relationship between the film and the substrate was determined through the use of X-ray reciprocal space maps. We characterized the ferroelectric domain structures of the films using conventional Piezoresponse Force Microscopy (PFM) in an area of 5 μm$^2$. All the films show no piezoresponse contrast in either the out-of-plane or the in-plane directions, confirming the ferroelectric monodomain structure at 10.4% Sm doping, and paraelectric phases at 16.1 and 18.7% Sm doping.

Neutron diffraction measurements were performed on: the BT-4 and BT-7 triple-axis spectrometers (NIST Center for Neutron Research); the IN20 thermal triple-axis spectrometer (ILL, France) and the cold triple-axis spectrometer TASP (PSI, Switzerland). The measurements on BT-7,[20] IN20 and TASP employed polarized neutrons and IN20 and TASP were equipped with the 'zero magnetic field' environments CRYOPAD[21] and MuPAD[22], respectively. The zero-field environments allow the arbitrary orientation of the neutron polarization, and as such all scattering cross-sections can be measured. Polarized neutron measurements allow a direct measurement of the orientation of a magnetic moment in the scattering plane and can be used to identify a magnetic structure that is composed of two out of phase components i.e. a spiral or cycloidal structure gives a different response to a modulated structure with the same periodicity. All measurements were done at room temperature.

Fig. 1 shows the X-ray reciprocal space maps (RSMs) taken for 10.4% and 18.7% Sm-doped BFO films. We see there is one broad reflection for Sm concentration less than or more than the MPB (~14% Sm). Therefore, we regard the films used in this study as consistent with a single crystallographic domain. For concentrations greater than the MPB, RSMs reveal that the material is orthorhombic.[10]

For the 10.4% Sm-doped BFO film, which is Sm concentration less than the MPB in the ferroelectric rhombohedral phase, we focused on the magnetic reflections observed near the (0.5 0.5 0.5) reciprocal lattice position (we use pseudocubic indexing throughout this paper). The reflection lies in the (H K (H+K)/2) scattering plane, which is defined by the (1 1 1) and (1 -1 0) reflections. Fig. 2(a) shows polarized diffraction measurements of Spin-Flip (SF) and Non-Spin-Flip (NSF) scattering with the neutron polarization parallel or anti-parallel to the scattering vector, Q. In this figure, we see that SF scattering (+- and -+) shows two peaks, while the NSF scattering shows no peak. Therefore, these peaks have an entirely magnetic origin. The slightly incommensurate peak positions (with respect to the film) indicate the presence of either a modulated or chiral magnetic structure. Reciprocal space maps made in this scattering plane indicate that the propagation direction is along (1 -1 0). The clear difference between +- and -+ scattering, observed for one of the peaks, excludes a modulated collinear structure. Fig. 2(b) shows SF and NSF scattering with the neutron polarization out of the scattering plane. In this figure, we see that the NSF scattering has two peaks, with weaker scattering in the SF channels. This evidences that the cycloid or skew-helix (elliptic or circular), when projected perpendicular to Q, has a significantly larger component along the out-of-plane direction (1 1 -2) than in the plane, along (1 -1 0). Nevertheless, this projected ellipticity alone cannot explain the extremely small difference in +- and -+ channels of Fig. 2(a), which appears to point to near-equally populated chirality domains.

Spherical Neutron Polarimetry (SNP) measurements undertaken on TASP with MuPAD could not fully resolve the details of the magnetic structure. However, the scattering shown in Fig.2 (measured on BT-7) and measured on TASP indicate that the magnetic structure has components parallel to (1 1 -2) as well as along (1 -1 0). Thus, while we cannot say with certainty whether the magnetic structure is a skew-helix, or elliptic cycloid, the data suggests the presence of some chirality in the structure, although with a spin-plane different from the one found in the undoped BFO.[4]

Next, we turn to the film with Sm concentration greater than the MPB, which is in the paraelectric orthorhombic phase. Fig.3 shows our polarized neutron diffraction measurements for the 16.1% Sm-doped BFO film. This time, we oriented the film in the (H H L) scattering plane, which is defined by the (1 1 0) and (0 0 1) reflections. We show SF and NSF measurements on the magnetic reflection around the (0.5 0.5 0.5) position with the neutron polarization P parallel to the scattering vector Q (Fig. 3(a)). A single peak is observed for the SF and NSF configurations, indicating that there is a (0.5 0.5 0.5) ordering wave vector present, which implies G-type antiferromagnetic order. When P is out of the scattering plane as shown in Fig. 3(b), the SF and NSF scattering also show a single peak. This result indicates that the magnetism has components along [1 -1 0] and [1 1 -2]. Since we are not able to restrict the moment in the HHL plane only from (0.5 0.5 0.5) reflection due to the insensitivity of magnetic neutron scattering to components of the moment along the scattering vector, [1 1 1], we also measured (0.5 0.5 -0.5) reflection to determine the moment as shown in Fig.3(c, d). SF scattering at P∥Q

(Fig.3(c)), NSF and SF scattering at P⊥Q (Fig.3(d)) show a single peak, indicating the moment has components along [–1 1 0] and [1 1 2]. We determine the integrated intensities of the magnetic scattering peaks from (0.5 0.5 0.5) and (0.5 0.5 -0.5) by fitting as shown in Fig. 3 after correcting for depolarization of the $^3$He cell (TABLE I). Then, the possible orientation of the magnetic moment has been extracted by fitting those integrated intensities (of all measured cross-sections) using a Bayesian fitting routine: the moment has the angle φ = 15.1(2)° from [1 0 0] and θ = 56.1(4)° from [0 0 1] direction.

We also measured the higher Sm-doped BFO film with 18.7% Sm doping using Cryopad at ILL. The fit to the polarization matrix data measured at the magnetic reflections (0.5 0.5 0.5) and (0.5 0.5 -0.5) was undertaken with MuFit,[23] and is shown in TABLE II. The magnetic structure of the film is also a G-type collinear structure. The best fit gave an orientation of the magnetic moment of φ = 31.4(4)° from [1 0 0] and θ = 43.8(2)° from [0 0 1]. While we initially attempted to fit our data to a model consisting of single domains, such fits were poor (even using simulated annealing) and we switched to a model allowing magnetic domains of unequal populations. As the film is orthorhombic in this phase, we allowed 4 domains. There is the original domain (abc→abc), one in which the direction along the crystallographic a-axis is inverted (a→-a), (b→-b), and (c→-c). We found the sample consisted of the following domains 6.5% (abc→abc), 2.8% (c→-c), 13.1% (b→-b), and 77.6% (a→-a). We assume that this particular distribution of domains is accidental, though the strong preference for "a→-a" type domains is intriguing.

In Fig. 4, we summarize the evolution of the possible magnetic structures with Sm doping in the relatively thick (~1 μm) BFO films grown on miscut (001) STO substrates across the MPB. For the non-doped BFO film, the recovery of the cycloidal structure with a different chirality axis from bulk was found as we reported previously.[4] The 10.4% Sm-doped BFO, which has a Sm concentration less than the MPB, has a chiral magnetic structure propagating along (1 -1 0) with a component along both [1 -1 0] and [1 1 -2] directions. The recovery of cycloid structure in the relatively thick films used in this study might be due to the relaxation of the film as shown in the previous reports.[3,4] With concentrations greater than the MPB, at 16.1 and 18.7% Sm doping, the magnetic structure changes to G-type collinear, however the moment lies along different directions at different doping. These results indicate that the magnetic structure in the Sm-doped BFO film is very sensitive to the doping concentrations. This may be due to the effect of the chemical pressure from the Sm substitution into Bi site in the BFO films. The recent study on epitaxial BFO thin films with thickness of 70 nm on different substrates suggested that the magnetic structure changes drastically with small epitaxial strains.[25] Depending on the strain, it was shown in Ref. 25 that the magnetic structure changes from a spin cycloid, to a G-type antiferromagnet. While the thickness and substrate are different from our study, this report supports our observations as do theoretical predictions,[26] which show that the cycloid is extremely sensitive to strain. Furthermore, with Sm doping, both the Fe-O-Fe bond angle, as well as the lattice parameters shift, which could serve to change the single ion anisotropy direction and thus the direction of the ordered moment that we observed in the G-type antiferromagnetic samples.

In summary, we find that the doping of Sm into BFO thin films has a dramatic effect on its magnetic structure. As the Sm concentration increases, the magnetic structure evolves from a cycloid to a G-type antiferromagnet. Even within the G-type phase, the direction of the ordered moment changes. At the MPB, an electric-field-induced transformation from a paraelectric orthorhombic phase to the ferroelectric rhombohedral phase was suggested in our previous report.[10] Thus, Sm doping would allow one to tune the magnetic structure of the paraelectric phase. This phase could then be driven with an electric field back to the rhombohedral ferroelectric phase with cycloidal magnetic order. This opens the door to greater control magnetolectric devices across the phase boundary.

Bayesian fitting of the polarized neutron diffraction data at NIST was done using BUMPS developed by Dr. Paul Kienzle (NIST). The work was also supported by the W. M. Keck Foundation.

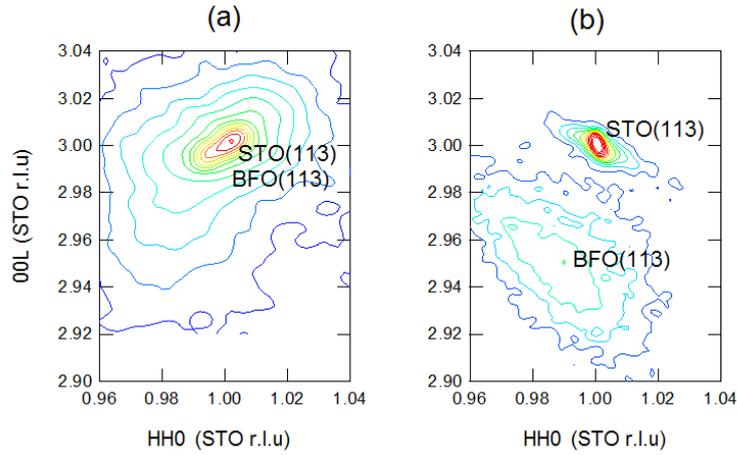

FIG. 1. X-ray reciprocal space mappings around STO (113) Bragg reflections taken for (a) 840 nm thick 10.4% Sm-doped BFO and (b) 1 μm thick 18.7% Sm-doped BFO films grown on miscut STO (001) substrate. Note that the peaks in 10.4% Sm-doped BFO are overlapping due to the small lattice mismatch.

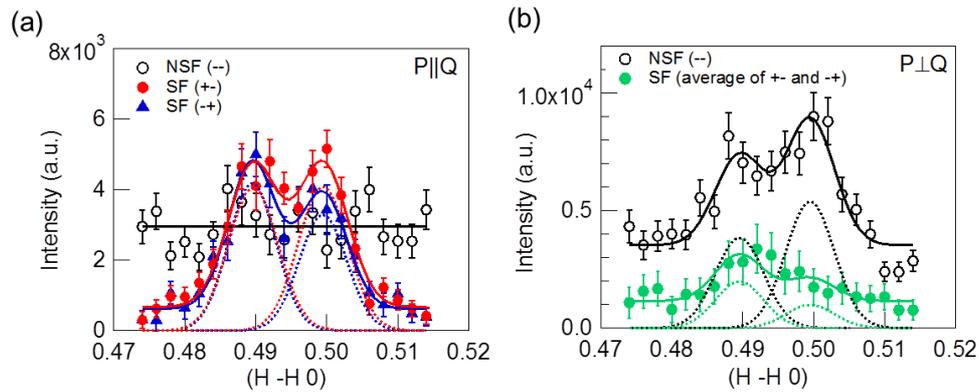

FIG. 2. Polarized neutron diffraction measurements of the 10.4% Sm-doped BFO film. Intensity profiles are taken in (H K (H+K/)2) zone with (a) P (neutron polarization) ∥ Q (scattering vector) and (b) P ⊥ Q. Open circles indicate Non-Spin-Flip (NSF), close circles and triangles indicate Spin-Flip (SF) scattering. Error bars are statistical in nature and represent one standard deviation. Solid lines represent fits to the data to two Gaussians and a background, dotted lines represent the Gaussian contribution to the fits.

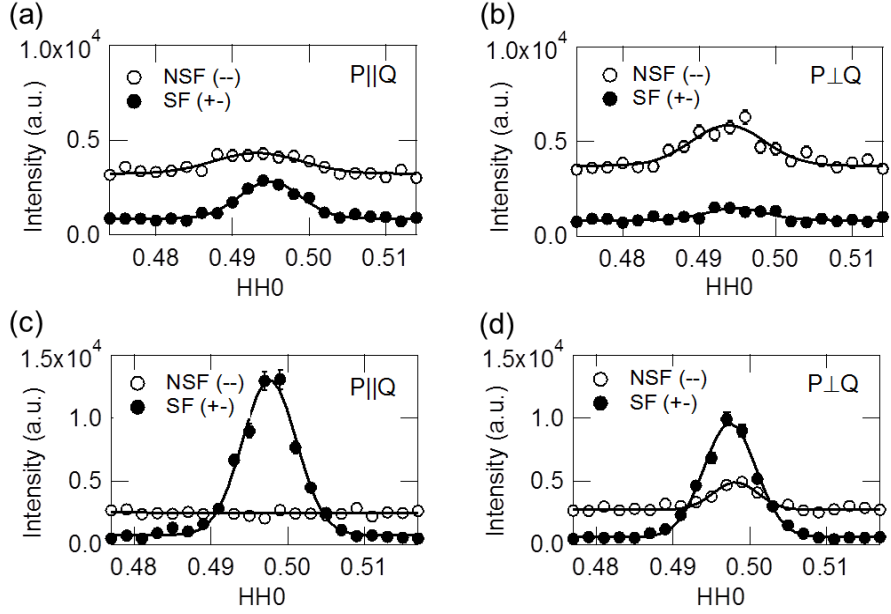

FIG. 3. Polarized neutron diffraction measurements of the 16.1% Sm-doped BFO film. Intensity profiles are taken in the (HHL) zone with (a) P∥Q and (b) P⊥Q about the (0.5 0.5 0.5), and (c) P∥Q and (d) P⊥Q about the (0.5 0.5 -0.5) position. Open circles indicate NSF, close circles and triangles indicate SF scattering. Error bars are statistical in nature and represent one standard deviation. Solid lines represent Gaussian fits to the data.

TABLE I. Integrated intensity of the peaks on the magnetic reflections (0.5 0.5 0.5) and (0.5 0.5 -0.5) of the 16.1% Sm-doped BFO film shown in FIG. 3.

| | P∥Q | | P⊥Q | |
|---|---|---|---|---|
| **Magnetic reflection** | NSF(--) | SF(+-) | NSF(--) | SF(+-) |
| (0.5 0.5 0.5) | 15.6±2.7 | 19.3±1.2 | 25.5±3.5 | 6.1±1.2 |
| (0.5 0.5 -0.5) | 0 | 108.4±4.1 | 15.8±1.1 | 76.3±2.3 |

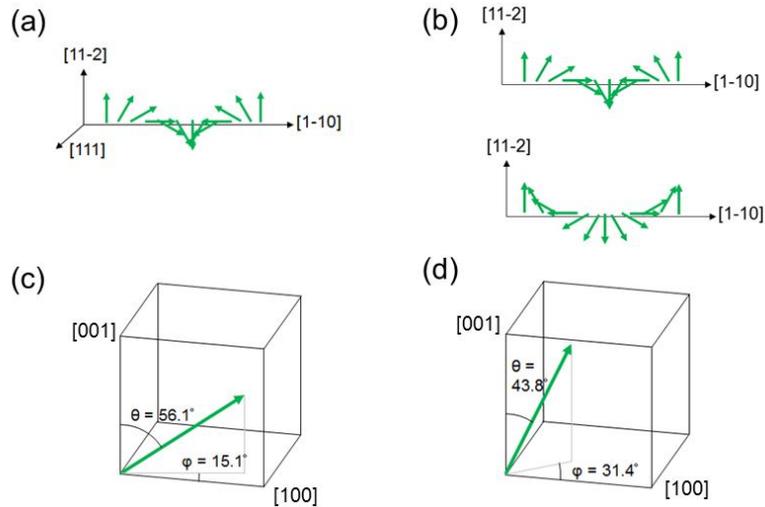

FIG. 4. Cartoon of possible magnetic structures for Sm-doped BFO films with Sm content of (a) 0% with a spin cycloid, (b) 10.4%, a cycloid with two chirality domains (c) 16.1%, a G-type antiferromagnet and (d) 18.7% a G-type antiferromagnet.

TABLE II. Polarization matrix data observed on the magnetic reflections (0.5 0.5 0.5) and (0.5 0.5 -0.5) of the 18.7% Sm-doped BFO film using Cryopad.

| Magnetic reflection | Pin | Pout (Measurement) | | | Pout (Calculated) | | |
| --- | --- | --- | --- | --- | --- | --- | --- |
| | | x | y | z | x | y | z |
| (0.5 0.5 -0.5) | x | -0.95(10) | 0.01(5) | 0.01(5) | -1 | 0 | 0 |
| | y | -0.02(5) | -0.07(7) | -0.52(8) | 0 | -0.14 | -0.52 |
| | z | 0.01(5) | -0.52(8) | 0.21(7) | 0 | -0.52 | 0.14 |
| (0.5 0.5 0.5) | x | -1.04(16) | -0.05(7) | -0.10(8) | -1 | 0 | 0 |
| | y | -0.08(8) | -0.1(1) | -0.76(14) | 0 | 0.01 | -0.75 |
| | z | 0.00(100) | -0.75(14) | -0.08(11) | 0 | -0.75 | 0.01 |